\documentclass[sigconf,review=False]{acmart}
\pdfoutput=1
\usepackage{graphicx}
\usepackage{epsfig}
\usepackage{color}
\usepackage{subfigure}
\usepackage{tabularx}

\usepackage{amssymb}
\usepackage{latexsym}
\usepackage{amsmath}
\usepackage{algorithm}
\usepackage[noend]{algpseudocode}
\usepackage{color}
\usepackage{relsize}
\usepackage[multiple]{footmisc}
\usepackage{mathtools}
\usepackage{bbold}
\usepackage{soul}
\usepackage{url}
\usepackage{blindtext}
\usepackage{multirow}

\AtBeginDocument{%
  \providecommand\BibTeX{{%
    \normalfont B\kern-0.5em{\scshape i\kern-0.25em b}\kern-0.8em\TeX}}}

\setcopyright{acmcopyright}
\copyrightyear{2024}
\acmYear{2024}
\acmDOI{10.1145/xxxxxxx.xxxxxxx}

\acmConference[WiSec '24]{WiSec '24: 17th ACM Conference on Security and Privacy in Wireless and Mobile Networks}{May 27--May 30, 2024}{Seoul, Korea}
\acmBooktitle{WiSec '24: 17th ACM Conference on Security and Privacy in Wireless and Mobile Networks, May 27--May 30, 2024, Seoul, Korea}
\acmPrice{15.00}
\acmISBN{978-1-4503-XXXX-X/21/06}

\begin{document}

\title{System-level Analysis of Adversarial Attacks and Defenses on Intelligence in O-RAN based Cellular Networks}

\author{Azuka Chiejina}
\affiliation{%
  \institution{\textit{NextG Wireless Lab}}
  \institution{George Mason University}
  \city{Fairfax}
  \country{USA}}
\email{achiejin@gmu.edu}

\author{Brian Kim}
\affiliation{%
  \institution{Northeastern University}
  \city{Boston}
  \country{USA}}
\email{br.kim@northeastern.edu}

\author{Kaushik Chowhdury}
\affiliation{%
  \institution{Northeastern University}
  \city{Boston}
  \country{USA}}
\email{krc@ece.neu.edu}

\author{Vijay K. Shah}
\affiliation{%
\institution{\textit{NextG Wireless Lab}}
  \institution{George Mason University}
  \city{Fairfax}
  \country{USA}
}
\email{vshah22@gmu.edu}

\renewcommand{\shortauthors}{Chiejina et al.}


\begin{abstract}
While the open architecture, open interfaces, and integration of intelligence within Open Radio Access Network technology hold the promise of transforming 5G and 6G networks, they also introduce cybersecurity vulnerabilities that hinder its widespread adoption. In this paper, we conduct a thorough system-level investigation of cyber threats, with a specific focus on machine learning (ML) intelligence components known as xApps within the O-RAN's near-real-time RAN Intelligent Controller (near-RT RIC) platform. Our study begins by developing a malicious xApp designed to execute adversarial attacks on two types of test data - spectrograms and key performance metrics (KPMs), stored in the RIC database within the near-RT RIC. To mitigate these threats, we utilize a distillation technique that involves training a teacher model at a high softmax temperature and transferring its knowledge to a student model trained at a lower softmax temperature, which is deployed as the robust ML model within xApp. We prototype an over-the-air LTE/5G O-RAN testbed to assess the impact of these attacks and the effectiveness of the distillation defense technique by leveraging an ML-based Interference Classification (InterClass) xApp as an example. We examine two versions of InterClass xApp under distinct scenarios, one based on Convolutional Neural Networks (CNNs) and another based on Deep Neural Networks (DNNs) using spectrograms and KPMs as input data respectively. Our findings reveal up to 100\% and 96.3\% degradation in the accuracy of both the CNN and DNN models respectively resulting in a significant decline in network performance under considered adversarial attacks. Under the strict latency constraints of the near-RT RIC closed control loop, our analysis shows that the distillation technique outperforms classical adversarial training by achieving an accuracy of up to 98.3\% for mitigating such attacks.

\end{abstract}




\maketitle

\section{Introduction}
\label{sec:intro}

Emerging applications, such as augmented and virtual reality (AR/VR), ultra-HD video streaming, autonomous vehicles, and industrial IoT, demand wireless systems capable of delivering superior performance in terms of high throughput, minimal latency, and high reliability. This has led to the development of new flexible, programmable radio access network (RAN) architectures. 
One such prominent RAN architecture gaining global traction is the Open Radio Access Network (O-RAN), developed by the O-RAN Alliance. O-RAN represents a groundbreaking approach that promises to usher in the next generation of cellular networks. These networks are envisioned to be open, interoperable, flexible, disaggregated, and intelligent, thanks to the O-RAN paradigm. O-RAN's architectural design aims to empower network operators to build adaptable and cost-effective networks that seamlessly accommodate a growing array of applications demanding high throughput, ultra-low latency, and expansive bandwidths.

The core O-RAN architectural principle encompasses the utilization of machine learning-driven closed-loop control, facilitated by RAN Intelligent Controllers (RICs) \cite{abdalla2022toward, 10024837}. There are two types of RIC, called, near-real-time (near-RT) RIC and non-real-time (non-RT) RIC that respectively support $3^{rd}$ party microservices, called xApps and rApps respectively. (Refer to Section 2 for an overview of O-RAN and RIC platform.) These microservices, equipped with a diverse array of machine learning (ML) techniques, are instrumental in various RAN applications such as scheduling, traffic steering, interference classification, and network slicing. The efficacy of these applications has been substantiated through extensive research and implementation in recent works \cite{bonati2021intelligence, upadhyaya2022prototyping, guillem2023SenseORAN,polese2022colo}.



Despite the many benefits that O-RAN brings{\def\thefootnote{}\footnotetext{This paper has been accepted for publication in ACM WiSec 2024}}, there is a growing concern regarding the vulnerability of various components hosted in the RICs and the vulnerability of the open interfaces~\cite{Groen23,ericcson, NTIA_ORAN_security_report, CISA_ORAN_security_report}. O-RAN Alliance Working Group 11 (Security Working Group) in \cite{O-RAN.WG11.Threat-Model.O-R003-v06.00} have done a comprehensive security analysis and identified various threat models that exist in O-RAN including threat agents, threat surfaces, and threats for each O-RAN component and open interfaces. Specifically, in \cite{O-RAN.WG11.Security-Near-RT-RIC-xApps-TR.0-R003-v03.00}, O-RAN Alliance Working Group 11 identified various attack vectors and threat models that could affect ML solutions hosted as xApps in the near-RT RIC. These threats could range from poisoning test data used by ML models to altering a ML model and breaching ML data confidentiality and privacy. Table \ref{tab:ML_intelligence_issues} provides a succinct summary of the various attack vectors that impact the application of ML intelligence within O-RAN, particularly concerning xApps and rApps. \textit{These findings underscore the critical importance of safeguarding intelligence and data deployed within the O-RAN system.} 

\begin{table*}[ht]
\caption{ML intelligence specific issues in xApps (as per O-RAN Alliance Security Working Group technical report on ``Study of Security for near-RT RIC and xApps''~\cite{ORAN_Security_Near_RT_RIC_xApp}).}
\vspace{-0.2in}
\scriptsize
\begin{center}
\begin{tabular}{|p{1.4in}|p{1.4in}|p{3.6in}|}
    \hline
      \textbf{Attack vector} & \textbf{Security threats} & \textbf{Consequence} \\ \hline
    \multirow{3}{1.4in}{ML model altercation, duplication or inversion} 
    & Malicious/Compromised xApp & - Malicious/compromised xApps misuse radio network data and control capabilities over RAN operations, disrupt subscriber services, take advantage of UE identification, locate UEs, and alter UE slice priority. \\ 
    &Conflicting xApp & - A denial of service (DoS) attack occurs when an xApp purposefully makes decisions regarding the management of radio resources that are in opposition to those made internally by the gNB or by other xApps. \\ 
    &Isolation between xApps & - Utilize too many of the platform resources that are shared by all xApps, causing a noisy neighbor effect by depleting the resources of other co-located xApps.\\ \hline

    \textit{ML training or test data poisoning (\textbf{Focus of this paper})} & \textit{Compromised ML data pipeline used in near-RT RIC} & \textit{Malicious attacker gains the ability to change some of the training (or test) data that may be used to develop the ML models that xApps will use to produce predictions for a certain RAN control. This would lead to predictions that will degrade network performances.} \\ \hline

    Privacy leaks in RAN database & Sensitive RAN and UE-specific data stored in near-RT RIC database  & Allows unauthorized database activities and information manipulation by an intruder. This could affect network service and performance. It also brings issues related to the privacy of users' data.\\ \hline
\end{tabular}
\vspace{-0.15in}
\end{center}
\label{tab:ML_intelligence_issues}
\end{table*}

So far, the mentioned studies have only laid out potential threats to O-RAN with no system-level study conducted to show the extent these threats can affect the performance of the system or the network in general. Threats against ML models to cause misclassification through carefully crafted perturbations have been studied in the field of adversarial machine learning \cite{szegedy2013intriguing, goodfellow2014explaining,kurakin2018adversarial}. With the growing application of ML in the wireless domain, adversarial attacks and threats have also been explored, particularly in the wireless physical layer \cite{sadeghi2018adversarial,sagduyu2019adversarial,shi2018adversarial,kim2021channel}. In \cite{sapavath2023experimental}, the authors gave the first and potentially the only known work on investigating adversarial attacks on xApps in O-RAN architecture. However, this work is limited in the sense that (i) the work does not explore any defense techniques to address the cyber vulnerabilities, (ii) the cellular network used in their study is based on MATLAB simulations, and does not reflect the system-level study, (iii) the work is limited to spectrograms, and has not explored KPMs which is mostly used by ML-based xApps in current O-RAN systems.

\textbf{Contributions.} In this paper, we undertake an in-depth examination of the cyber vulnerabilities inherent in the intelligent components of the O-RAN framework, specifically focusing on the ML models deployed within xApps at the near-RT RIC, considering a systemic perspective. We illustrate our investigation using a representative interference classification xApp, designed to detect whether the network is experiencing interference or not. It's worth noting that while we utilize this particular xApp as a case study, our analysis is applicable to any xApp employing predictive models for network control. Our specific area of concern centers around one of the threat vectors highlighted in Table \ref{tab:ML_intelligence_issues}, where we manipulate the test data employed by ML models within xApps in the near-RT RIC. We delve into the repercussions of these threats on network performance and subsequently explore techniques to counteract such attacks in real time, ultimately aiming to maintain robust network performance even in the presence of adversarial activities. In summary, this paper makes the following key contributions.

\smallskip \noindent $\bullet$ We conduct a comprehensive system-level investigation of cybersecurity threats, with a specific focus on adversarial attacks on the intelligent components, called xApps hosted within the near-RT RIC of O-RAN systems.

\vspace{-0.05in}
\smallskip \noindent $\bullet$ We create a malicious xApp designed for the execution of adversarial attacks on two different types of test data --  spectrograms and key performance metrics (KPMs), stored in the RIC database in the near-RT RIC. For our study, we leverage a ML-based interference classification (InterClass) xApp as the case study. For spectrograms as input data, we utilize a CNN model for the InterClass xApp (InterClass-Spec xApp), whereas we utilize a DNN model for InterClass xApp (InterClass-KPM xApp) for KPM as input data. 

\vspace{-0.05in}
\smallskip \noindent $\bullet$ We leverage the distillation technique as a defense to mitigate the cyber threats against intelligent components within the near-RT RIC. This technique involves training a teacher model at a high softmax temperature and transferring its knowledge to a student model trained at a lower softmax temperature. The robust model is eventually deployed within the xApp.

\vspace{-0.05in}
\smallskip \noindent $\bullet$ We employ an LTE/5G O-RAN testbed for the generation of datasets and assessing the impact of adversarial attacks. Furthermore, we evaluate the effectiveness of defense techniques on both model accuracy and network performance. Under the strict latency requirements of between 10ms to 1s for the near-RT RIC closed control loop, our experiments demonstrate that both the InterClass-Spec xApp and InterClass-KPM xApp models perform admirably, yielding accuracy rates of $\textbf{98\%}$ and $\textbf{97.9\%}$, respectively. However, when subjected to an adversarial attack with an epsilon value of \textbf{0.1} which is a hyperparameter that determines the magnitude of perturbation added to the input data to generate an adversarial example, we observe a substantial reduction in the accuracy of both models. Specifically, the InterClass-Spec model's accuracy drops to $\textbf{0\%}$, and the InterClass-KPM xApp model's accuracy decreases to $\textbf{3.7\%}$, resulting in network performance degradation. Notably, by employing the distillation technique as a defense mechanism, we managed to restore the accuracy of both models. The InterClass-Spec xApp model achieves an accuracy of $\textbf{96\%}$, while the InterClass-KPM xApp model achieves an accuracy of $\textbf{98.3\%}$ thereby improving the overall network performances in terms of throughput and BLER. 
\vspace{-0.05in}
\section{Related Works}

A key pillar of O-RAN architecture is the integration of intelligence into every aspects of wireless networks, whether its deployment, operation or maintenance of the networks. 
Recent years have seen an emergence of research papers that propose innovative AI-powered RAN control functionalities. Notable examples include: (i) \textit{LSTM-based RAN resource management xApp} \cite{niknam2022intelligent} - this work is potentially the first work that proposes utilizing long short-term memory (LSTM) recurrent neural network (RNN) to learn and predict the traffic pattern of a real-world cellular network using O-RAN networks - , (ii) \textit{ML-based spectrum sensing xApp}~\cite{guillem2023SenseORAN} - this work proposes ML-based spectrum sensing xApp that utilizes an object detection ML model, called, YOLO, for detecting radar signals present within the spectrograms in uplink LTE/5G communications, (iii) \textit{AI-driven traffic steering xApp}~\cite{lacava2023programmable} - this work proposes a deep reinforcement learning (DRL) based traffic steering xApp to optimally control mobility procedures at a UE level, using the centralized viewpoint of the RIC (using collected RAN KPMs), (iv)  \textit{DRL-based RAN slicing}~\cite{bonati2021intelligence} - this work develops a set of DRL agents as RIC xApps to optimize key performance metrics for different network slices through data-driven closed-control loops. These are just a few examples of the innovative research in this domain. There are numerous other works, such as DRL-based scheduling xApps~\cite{kouchaki2022actor}, ML-based mobility management solutions~\cite{prananto2022ran}, ML-based interference mitigation technique~\cite{10211189}, and ML strategies for traffic steering xApp~\cite{tamim2023intelligent}. This surge in research reflects the growing interest and advancements within the O-RAN ecosystem. 

Nonetheless, despite this impressive array of ML-based RAN control contributions, a critical gap remains unaddressed. None of the existing endeavors delve into the intricate terrain of potential cyber threats and privacy vulnerabilities posed by ML-driven x/rApps, or the overarching AI/ML-infused closed-loop systems within the O-RAN networks. Specifically in \cite{O-RAN.WG11.Security-Near-RT-RIC-xApps-TR.0-R003-v03.00}, O-RAN alliance working group 11 identified some key ML-related security issues in the near-RT RIC architecture, relevant interfaces, xApps and application programmable interfaces (APIs). 
Authors in \cite{liyanage2023open} provides a  comprehensive review of the security and privacy risks associated with the O-RAN architecture. They discuss potential solutions, standardization efforts, and unique solutions based on blockchain and AI. They offer a comprehensive identification of the threats associated with the O-RAN architecture but with no comprehensive system-level analysis and validation on how some of these attacks can be implemented in the O-RAN architecture.

\vspace*{-0.05in}
\section{O-RAN Background}
\label{sec:system}
\begin{figure}
    \centering  \includegraphics[width=0.85\linewidth]{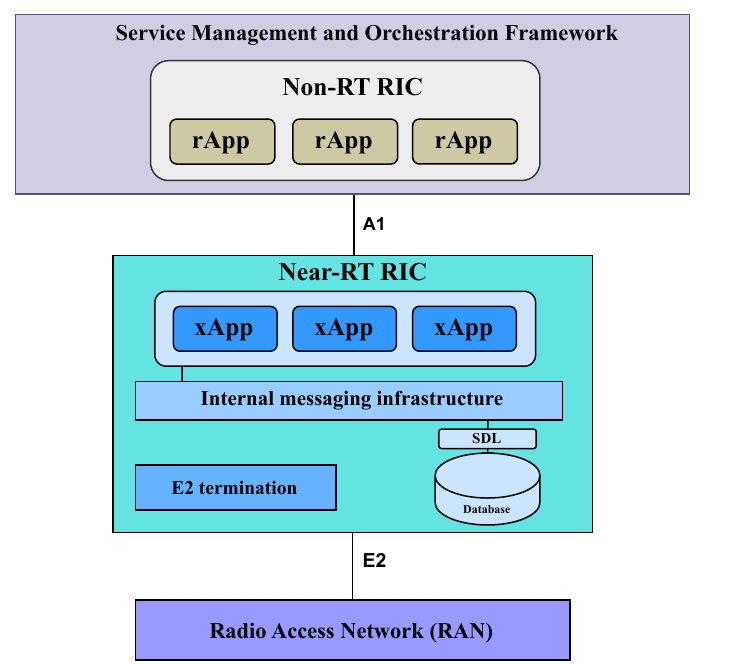}
    \vspace{-0.1in}
    \caption{Simplified O-RAN architecture.}
    \label{fig:oran}
    \vspace{-0.2in}
\end{figure}

The O-RAN architecture, as depicted in  Fig. \ref{fig:oran}, provides an overview of the internal components within an O-RAN system. Here, we briefly discuss three integral internal components for better understanding of this work. (Please refer to \cite{10024837} for a detailed understanding of O-RAN architecture.) 
\vspace{-0.15in}
\subsection{RAN Intelligent Controller (RIC)}
The RIC is a key component of the O-RAN architecture that processes data and leverage ML algorithms to determine control policies that optimizes the RAN. There are two major logical RIC controllers that operate different time scales. These are:

    \smallskip \noindent $\bullet$ \textbf{Non-RT RIC}: The non-RT RIC resides within the service management and orchestration framework (SMO) and handles control loops at a time granularity of $>1$s. 

    \smallskip \noindent $\bullet$ \textbf{Near-RT RIC:} The near-RT RIC operates control loops between 10ms to 1s. It hosts third-party vendor applications called \textbf{xApps}. These xApps act as \textit{intelligent components} and run ML algorithms that are used to determine control policies for optimizing the RAN through the E2 interface. Other major components of the near-RT RIC include the RIC database/SDL and internal messaging infrastructure which helps to connect multiple xApps and also ensures message routing. \textit{This work focuses on security analysis of ML model intelligence within xApps hosted in the near-RT RIC.} 

\vspace{-0.1in}
\subsection{RIC Database and Shared Data Layer (SDL)} 
\label{dbsdl}
The RIC database serves as a repository for various data, including lists of User Equipments (UEs) and associated information. It also contains data related to the RAN, offering insights into access network-related aspects influencing overall network performance. The data stored in the RIC database may encompass KPMs such as throughput and signal-to-interference-plus-noise ratio (SINR) characterizing the quality of communication between UEs and RAN.

The SDL, on the other hand, functions as an API enabling xApps to access and manipulate the information stored in the RIC database. Through the SDL, xApps have the ability to read, write, and modify data stored in the database. The O-RAN software community (OSC) in \cite{sdl}, provides documentation on the implementation of the SDL API, which can be compiled within xApps, granting them access to a Redis-based database.
\vspace{-0.15in}
\subsection{Types of Data within RIC Database}
Based on the split option used in the 5G network architecture, there is a corresponding data type that can be obtained from the lower network levels to the near-RT RIC.

\paragraph{\textbf{Key Performance Metrics (KPMs):}}
In line with the O-RAN alliance's split option 7.2x, which is specifically designed for Ultra Reliable Low Latency Communication (URLLC) and near-edge deployments, we need to take a crucial architectural consideration into account. Split 7.2x effectively relocates the low-level Physical Layer (PHY) functionality to the Radio Unit (RU). As a result of this architectural shift, we encounter a limitation in terms of the data we can leverage within the near-RT RIC.

In this scenario, our data sources are primarily constrained to KPMs. These metrics provide valuable insights into the performance of the network, which is especially relevant for URLLC and near-edge use cases. While this architectural change streamlines the processing of PHY-related data at the RU, it also underscores the significance of KPMs as the primary data source for informed decision-making within the near-RT RIC.
\vspace{-0.05in}
\paragraph{\textbf{I/Q Samples (or Spectrograms):}}
Conversely, under the split 8 configuration, the low-level PHY functionality remains within the Distributed Unit (DU-low). This architectural choice provides distinct advantages, particularly for data accessibility. With split 8, it becomes feasible to access a more extensive range of data types, extending beyond KPMs.

Certainly, the inclusion of IQ samples/spectrogram data in split 8's accessible data spectrum is a significant advancement. This expanded access to a diverse array of data types, even in the face of potential latency and security considerations, holds immense importance. It provides a more comprehensive dataset that proves invaluable for in-depth analysis and informed decision-making.

Therefore, by preserving the low-level PHY functions within the DU-low, split 8 effectively amplifies the network's adaptability and the variety of data sources at its disposal. This heightened versatility empowers the deployment of advanced use cases and applications, underscoring the potential of this configuration to drive innovation and address complex network requirements.

\section{Exemplary Interference Classification xApp}\label{sec:interfxapp}
To study the threat and vulnerability of intelligent components in the near-RT RIC, we design, develop, and deploy a ML-based InterClass xApp that is used to detect a jammer transmitting an interference signal in the uplink direction of the user traffic. Our models used for the xApp are trained on our dataset collected over the air (OTA) using our LTE/5G O-RAN testbed detailed later in Section \ref{sec:testbed}.

\begin{figure}
    \centering  \includegraphics[width=0.95\linewidth]{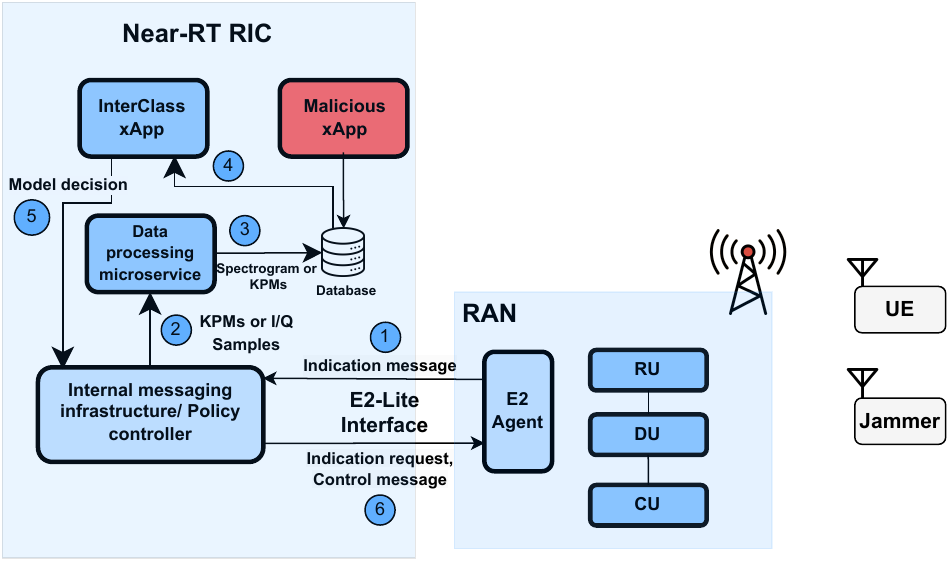}
    \vspace{-0.2in}
    \caption{Overview of Interference Classification xApp (with Malicious xApp). } 
    \label{fig:system}
    \vspace{-0.25in}
\end{figure}
\vspace*{-0.1in}
\subsection{Model Design and Development}
\label{modeldesign}
In the design and development of our models utilized within the InterClass-Spec xApp and InterClass-KPM xApp which use spectrogram and KPMs respectively, we encounter a unique challenge. These models need to meet the stringent demands of the near-RT RIC control loop, which operates at a timescale of less than a second. To ensure seamless operation within this timeframe, it is imperative that our models are not overly complex and burdened with an excessive number of parameters. Such complexity can hinder the speed of inference, which is a critical consideration in this context.

Hence, we find ourselves at a critical juncture where we must carefully balance model accuracy with the efficiency of inference speed. This tradeoff is essential to strike the right equilibrium, ensuring that our models deliver both the required accuracy and the rapid decision-making capabilities necessary for the near-RT RIC control loop.
\vspace{-0.05in}
\subsubsection{CNN-based InterClass-Spec xApp}
For the spectrogram dataset, we leverage the CNN for training purposes to build our initial model with input of shape (128,128,1) and having total parameters of 163,922. Table \ref{spec-model} shows our model structure where the first two conv2D layers have a maxpooling layer after it with pool size of (2,2) while a flatten layer is introduced after the last conv2D layer. 
\begin{table}
    \centering
    \caption{Spectrogram Dataset Model structure}
    \vspace{-0.1in}
    \scriptsize
    \begin{tabular}{|p{0.6in}|p{2.5in}|}
        \hline
         \textbf{Layers} & \textbf{Properties}  \\ \hline
        Conv2D & filters = 16, kernel size = (3,3), activation function = ReLU  \\\hline
        Maxpool2D & pool size = (2,2) \\\hline
         Conv2D & filters = 16, kernel size = (3,3), activation function = ReLU  \\ \hline
          Maxpool2D & pool size = (2,2) \\ \hline
         Conv2D & filters = 32, kernel size = (3,3), activation = ReLU  \\ \hline
         Maxpool2D & pool size = (2,2) \\ \hline
         Conv2D & filters = 32, kernel size = (3,3), activation = ReLU  \\ \hline
         Flatten & - \\ \hline
        Dense & size = 32, activation = ReLU  \\ \hline
          Dense & size = 2, activation = Softmax  \\ \hline
    \end{tabular}
    \label{spec-model}
    \vspace{-0.2in}
\end{table}
\vspace{-0.05in}
\subsubsection{DNN-based InterClass-KPMs xApp}
\begin{figure}
    \centering  \includegraphics[width=0.85\linewidth]{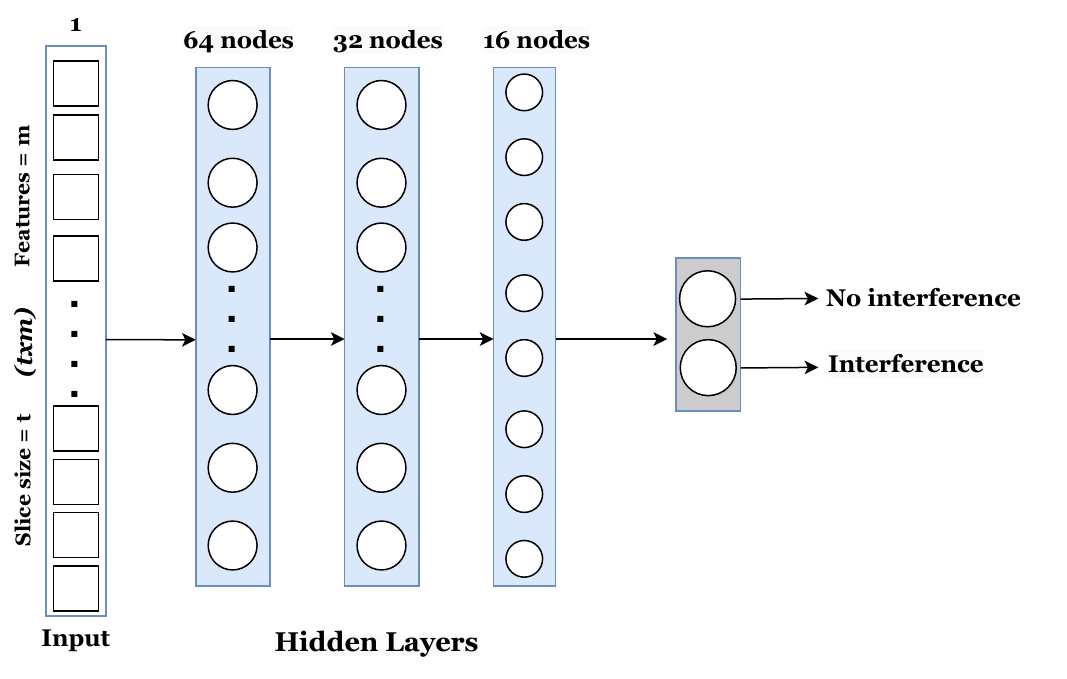}
    \vspace{-0.2in}
    \caption{KPMs model architecture.}
    \label{fig:kpm-model}
    \vspace{-0.2in}
\end{figure}

For the KPMs dataset, we leverage a DNN architecture as shown in Fig.\ref{fig:kpm-model} having a total of 6546 parameters. To train the DNN-based model, we leverage the metric values in the dataset which represent the features. In our case, we consider (m) KPMs and collect (t) different time windows of metrics to stack together and form an extended array of input before feeding into the network. Feeding multiple time windows into the model leads to better accuracy for interference classification compared to feeding a single time window to the model. For our training purposes, we considered m=4 and t=15, which means our input nodes would be 60 (15x4) in total resulting in an input shape of (60,1). Each hidden layer has a ReLU activation function while the final output has a softmax activation function. 

Upon model creation, we utilize the model in the InterClass xApp which is deployed in the near-RT RIC of our O-RAN system as shown in Fig.\ref{fig:system}. 

\vspace*{-0.1in}
\subsection{InterClass-Spec xApp - A Walkthrough}

We discuss the step-wise working of InterClass-Spec xApp. Also refer to Fig. \ref{fig:system} for illustration.

\noindent \textbf{Step 1)} I/Q samples collected OTA are stored in a buffer at the RAN during communication. After an E2 request is sent from the internal messaging infrastructure (IMI)/policy controller, a connection is established between the near-RT RIC components and the RAN. The request from the policy controller is in the form of an indication message to request for the I/Q samples stored in the buffer at the RAN. These I/Q samples are then collected through the E2 interface from the RAN to the policy controller. For our experimental purposes, we periodically collect the last $10$ ms segments of I/Q data from the buffer, equivalent to the length of one LTE/5G frame.

\noindent \textbf{Step 2)} The last $10$ ms I/Q samples are forwarded to an auxilliary \textit{data processing microservice} that processes and converts the I/Q samples into spectrograms.

\noindent \textbf{Step 3)} The \textit{data processing microservice} then forwards the computed spectrogram to a database that stores computed spectrograms that can be accessed by other xApps deployed in the near-RT RIC that require the data for different RAN control functionalities.

\noindent \textbf{Step 4)} The \textit{InterClass-Spec xApp} queries the database to get the latest spectrogram and uses it to determine if a signal is interfered or not.

\noindent \textbf{Step 5)} The decision/prediction of \textit{InterClass-Spec xApp} is sent to the policy controller. 


\noindent \textbf{Step 6)} The policy controller sends a control message to the RAN. The control message contains the control decision made by the xApp based on the model's decision. Based on this decision, we dynamically control the RAN. The RAN can be controlled to either use the highest achievable MCS if there is no jammer detected but if a jammer is detected, there are several mitigation approaches that could be implemented such as adaptive MCS, change of carrier frequency or perform some intelligent scheduling approaches based on resource blocks assignments based on the jamming signal detected. In our case, we use the adaptive MCS technique as the mitigation approach for negating the impact of interference on the network performance.


\subsection{InterClass-KPM xApp - A Walkthrough}

InterClass-KPM xApp works similarly to that of InterClass-Spec xApp with the only following differences. 

In \textbf{Step 1}, unlike I/Q samples, here the requested KPMs are sent as the indication request message to the policy controller when requested. In \textbf{Step 2}, KPMs are forwarded to the data processing microservice to perform certain processing before it is stored in the RIC database. After this, \textbf{Step 3-6} work in a similar fashion however using KPMs as the test data instead of spectrograms.

\section{O-RAN Intelligence Threat Model} 
In our O-RAN system, as illustrated in Fig.~\ref{fig:system}, our focus within the threat model centers on the O-RAN intelligence components, xApps operating within the near-RT RIC. Specifically, our research delves into potential threats targeting the test data employed by ML models within these xApps for inference, and the potential repercussions on network performance.

In the broader O-RAN context, ML models are integrated into xApps to facilitate the management of the RAN within a time range of 10ms to 1000ms, relying on the decisions made by these models. Despite the specified time scale for control, these models are susceptible to generating inaccurate predictions when faced with adversarial attacks that aim to manipulate the data used by the models. Numerous xApps can be developed within the near-RT RIC, each serving different use cases such as traffic steering, slicing, resource allocation, and interference classification. The models employed in these scenarios may necessitate data presented in formats like KPMs or spectrograms. The accuracy and efficiency of these models can be compromised by adversarial attacks during the testing phase. The impact of these attacks on network performance varies depending on the specific scenario. For instance, in the case of an xApp designed for slicing and resource allocation, incorrect predictions from the models could result in subpar quality of service and decreased data rates for users requiring additional resources.

For our study, we have implemented an ML-based InterClass xApp designed to detect the presence of jammers in the network. This xApp utilizes either spectrograms or KPMs data stored in the RIC database. Simultaneously, we've created a \textit{malicious xApp} that coexists within the shared near-RT RIC, assuming access to the same database due to the openness of O-RAN system. This malicious xApp can be crafted using well-known adversarial attack algorithms to manipulate the data stored in the shared database either during test time or in real-time. The objective of this malicious xApp in this scenario is to ensure that the data used for inference by the legitimate xApp during test time is altered. The intention is to compel the legitimate xApp to make incorrect predictions, thereby degrading network performance. The network performance metrics considered include throughput and block error rate (BLER)

Similar to the work in \cite{sapavath2023experimental}, for the malicious xApp, we leverage two well known adversarial attacks called (i) the \textbf{fast gradient sign method (FGSM) attack} \cite{kurakin2016adversarial} and (ii) the \textbf{projected gradient descent (PGD) attack} \cite{madry2017towards}.
\paragraph{\textbf{FGSM attack}}
We first leverage the FGSM to create adversarial samples to perform targeted attacks. FGSM is a one-step gradient-based attack where the adversarial attack $\boldsymbol{x}_{adv}$ is generated to minimize the loss function $\mathcal{L}(\boldsymbol{\theta},\boldsymbol{x},y^{target})$ where $\boldsymbol{\theta}$ is the model's parameters, $\boldsymbol{x}$ is the input to the model, and $y^{target}$ is the target label that the adversary is aiming to fool. In our case, since we are performing a targeted attack the adversarial xApp aims to fool InterClass xApp to classify as signal of interest (SOI) even in the presence of jammer signals. To minimize the loss function, the adversarial attack can be expressed as $ \boldsymbol{\delta} = \epsilon * \text{sign}(\nabla_{\boldsymbol{x}}\mathcal{L}(\boldsymbol{\theta},\boldsymbol{x},y^{target}))$, where $\nabla_{\boldsymbol{x}}\mathcal{L}(\boldsymbol{\theta},\boldsymbol{x},y^{target})$ is the gradient of $\mathcal{L}(\boldsymbol{\theta},\boldsymbol{x},y^{target})$ with respect to the input data $\boldsymbol{x}$ and $\epsilon$ is the power scaling factor of the attack which controls how far the input data can be modified while still remaining similar to the original data. The adversarial example that is put to the database is represented as $\boldsymbol{x}_{adv} = \boldsymbol{x} + \boldsymbol{\delta}$.
\vspace{-0.1in}
\paragraph{\textbf{PGD attack}}
The FGSM can be further improved by running a more thorough optimization using an iterative algorithm where this multi-step iterative algorithm based on FGSM is called PGD attack \cite{madry2017towards}. PGD attack performs FGSM with a small step size $\alpha$ and projects the perturbed input onto the $L_{\infty}$-ball around the original input. The iterative method for $N$-step PGD attack is defined for $n$-th iteration as follows
\vspace{-0.05in}
\begin{align}\label{eq:pgd attack}
    &\boldsymbol{x}_{0} = \boldsymbol{x} \nonumber\\
    &\boldsymbol{x}_{n} = \text{clip}_{[\boldsymbol{x},\epsilon]}\{\boldsymbol{x}_{n-1}+\alpha \text{sign}(\nabla_{\boldsymbol{x}_{n-1}}\mathcal{L}(\boldsymbol{\theta},\boldsymbol{x}_{n-1},\boldsymbol{y}))\} \nonumber\\ 
    &\boldsymbol{x}_{adv} = \boldsymbol{x}_{N},
\end{align}
where $\text{clip}_{[\boldsymbol{x},\epsilon]}(\cdot)$ is elementwise clipping to $[\boldsymbol{x}-\epsilon,\boldsymbol{x}+\epsilon]$ so that the outcome stays in $L_{\infty} \epsilon-$neighborhood of $\boldsymbol{x}$. 

In our threat model, we consider a \textit{white-box attack} scenario in which the adversary possesses unrestricted access and complete knowledge of the model. This approach is highly practical and aligns with the vulnerabilities and openness inherent in the O-RAN architecture, as discussed in Section \ref{sec:intro}.

In this scenario, the attacker, manifesting as an adversarial xApp, not only has access to the database but also acquires detailed knowledge about the model employed by the legitimate xApp. Leveraging the adversarial attack methods previously outlined, the attacker carries out targeted attacks in real-time. \textit{The primary objective is to manipulate the legitimate xApp's behavior, compelling it to classify a specific class erroneously. This is achieved by modifying the original data within the database, deceiving the legitimate xApp into making incorrect inferences.}

\subsection{Impact of Adversarial Attacks on InterClass-Spec/Interclass-KPMs xApp}
\begin{figure*}
\begin{center}
\subfigure[]{\label{fgsm-pgdmotivation(spec)}
\epsfig{figure=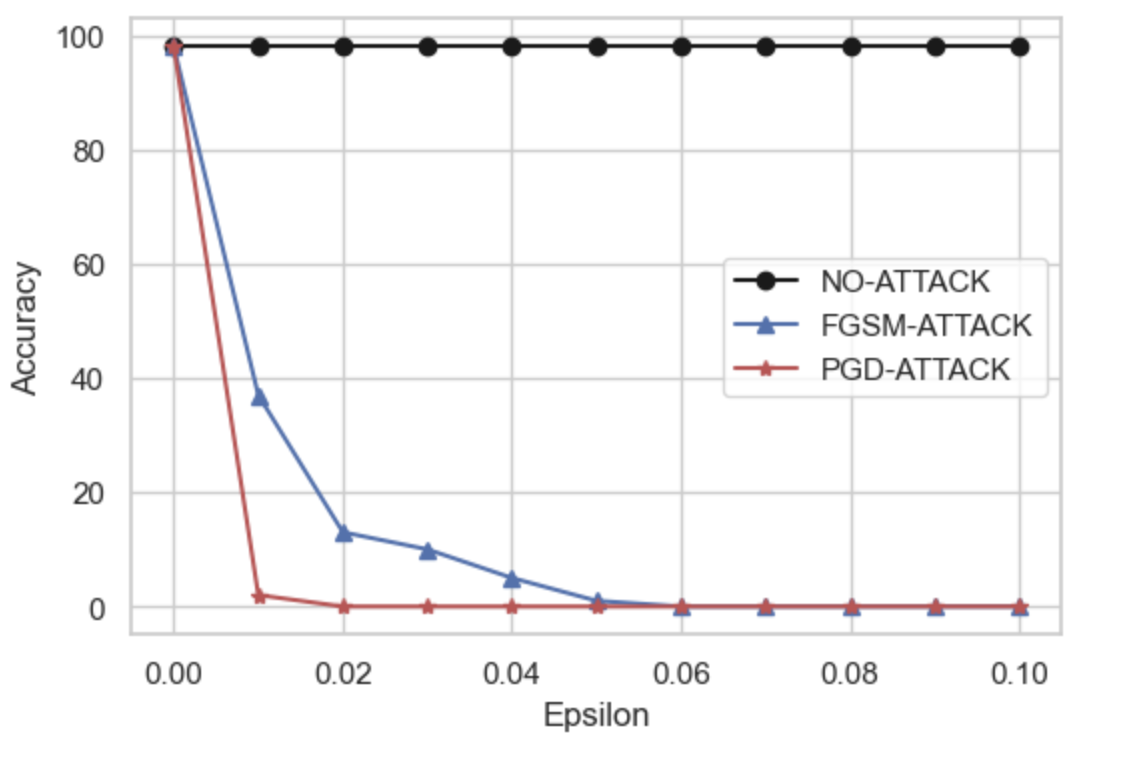,height=1.36in}}
\subfigure[]{\label{fgsm-pgdmotivation(kpms)}
\epsfig{figure=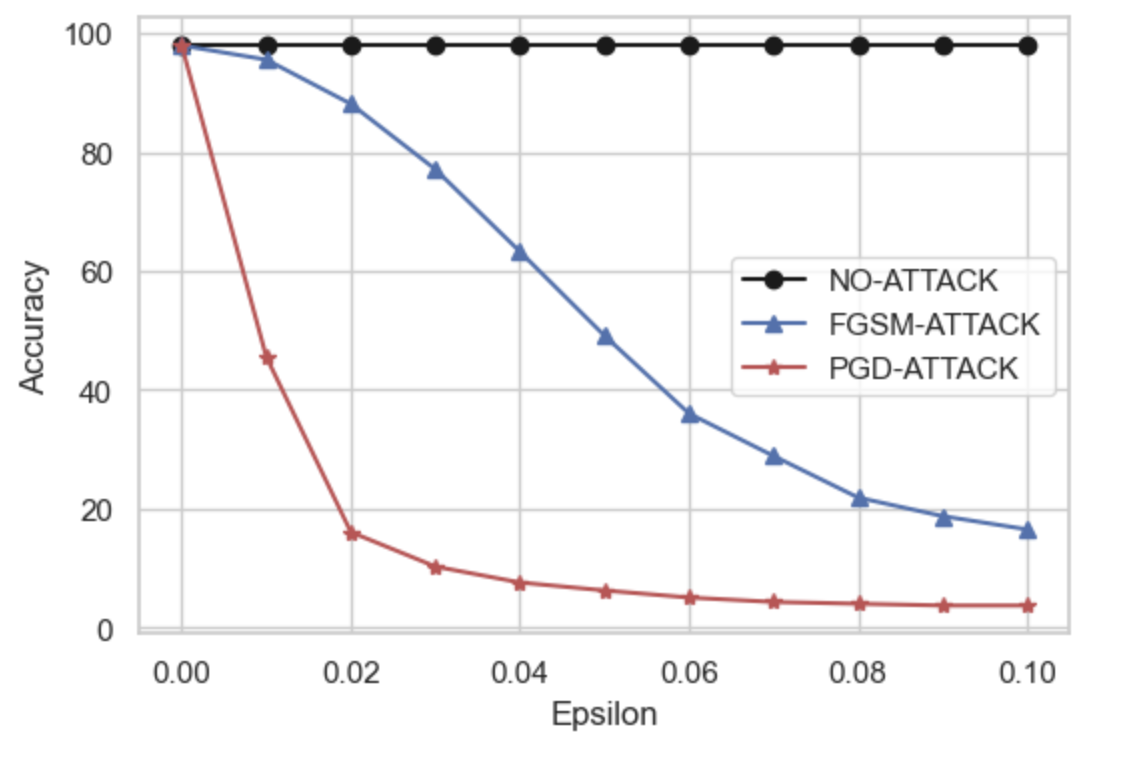, height=1.36 in}}
\subfigure[]{\label{spec-kpms}
\epsfig{figure=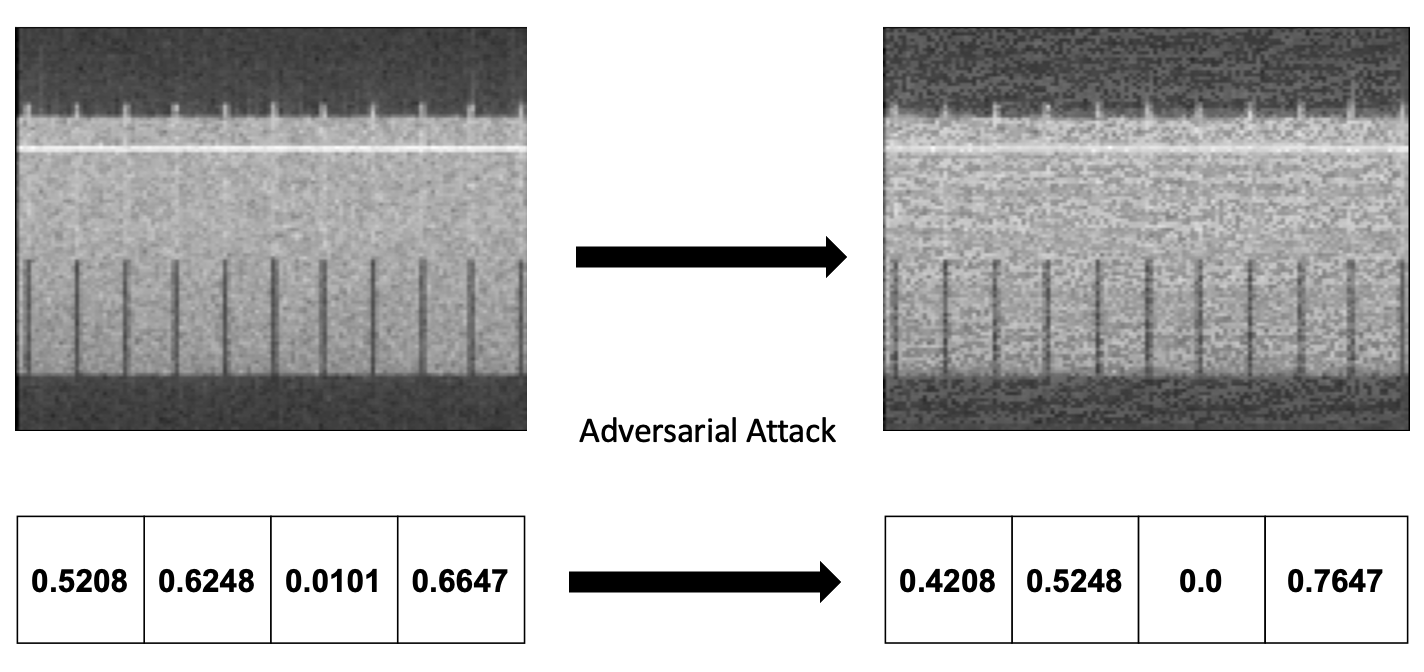, height=1.35 in}}
\vspace{-0.2in}
\caption{Impact of attacks on the models' performances -- (a) InterClass-Spec xApp accuracy, and (b) InterClass-KPMs xApp before and after the two attacks at various epsilon values, and (c) Example of original and perturbed spectrograms/KPMs under adversarial attack at $\epsilon=0.03$.}   \label{eps-motivation}
\end{center}
\vspace{-0.2in}
\end{figure*}

Following a comprehensive examination of our system-level threat model, we have obtained initial results indicating that the attacks on our models resulted in a degradation of network performance. Prior to these attacks, the accuracy of our models, one built with spectrograms and the other with KPMs, stood at an impressive $\textit{\textbf{98\%}}$ and $\textit{\textbf{97.9\%}}$, respectively. However, under the described attacks with varying perturbation budgets $\epsilon$ ranging from 0.01 to 0.1, with the FGSM attack being one-step while the PGD being five-step, the accuracy performances of our models are severely impacted and are depicted in Fig. \ref{eps-motivation}.

Fig. \ref{spec-kpms} present spectrograms and KPMs before and after the attacks at a perturbation budget of 0.03. The altered spectrogram exhibits clear visual differences, discernible to the human eye. However, these distinctions may not be apparent to the vulnerable model. Likewise, in the case of KPMs, we observe the change in KPM values resulting under adversarial attacks which cause misclassification. 

\section{Distillation-based Defense}
\label{defenses}
Drawing upon a state-of-the-art defense paradigm, namely the distillation-based defense, we present our approach to safeguard the integrity of intelligent components, specifically the xApps, against adversarial intrusions in the O-RAN system. This method forms a pivotal component of our comprehensive security framework, enhancing the resilience of O-RAN infrastructure while upholding the integrity and reliability of network functionalities.


Distillation was formally proposed by Hinton et al \cite{hinton2015distilling}. Distilling the knowledge in a neural network is a technique that helps to improve model efficiency, generalization, and regularization by reducing the size of neural network architecture thereby reducing the computing resources needed. In knowledge distillation, we have two neural networks architectures, one is called the teacher and the other is the student. The teacher is usually a larger architecture with more parameters and complexity while the student is usually of a smaller architecture. The motivation behind this is that distillation extracts the knowledge from the probability vectors of the teacher model and transfers it to the student model during training thereby maintaining comparable accuracy and improving generalization capabilities. In \cite{7546524}, Papernot \textit{et al.} proposed using the distillation technique not just for the transfer of knowledge between architectures, but also to make a model more resilient to adversarial examples. Their intuition was that this would also help in improving the generalization capabilities of the architecture while maintaining accuracy and increasing resilience to adversarial examples. A key difference between the approach of \cite{hinton2015distilling} and \cite{7546524} is that in using distillation as a defense technique, the size of the original/teacher architecture and the distilled architecture for training the student are kept the same. The student and teacher model architectures are kept the same in order to ensure that the knowledge transfer from the teacher to the student is effective and the student model can learn from the teacher's predictions and mimic its behavior more accurately. This helps in improving the generalization capabilities of the student model and reducing the impact of adversarial samples. This method has also been applied in the wireless domain \cite{ozgur2022defensive} where it was used to defend against attacks against channel estimation models.

For our studies, aside from keeping both teacher and student models the same, we first trained the teacher model with a high temperature ($T_t$) to soften the softmax probability outputs of the model, which makes the probability distribution of the generated softmax function more uniform. Softmax function can be shown as
\begin{equation}
P(i,T) = \frac{e^{z_i/T}}{\sum_{j=1}^{K} e^{z_j/T}}  \ \ \ \text{for}\ i=1,2,\dots,K
\label{softmax}
\end{equation}
where $P(i,T)$ is the softmax probability output for the $i$th class scaled with temperature parameter (scaling factor) $T$, $z_{\textit{{i}}}$ is the unormalized logit for the $i$th class, and $K$ is the number of classes.
The loss function for the teacher model with temperature $T_t$ becomes
\vspace{-0.1in}
\[\mathcal{L}_{\text{teacher}}(T_t) = -\frac{1}{N} \sum_{j=1}^{N}  \mathbb{1}(y_{\text{i}}=j) P(i,T_t)\]
\vspace{-0.2in}
\begin{equation}
 = -\frac{1}{N} \sum_{j=1}^{N}  \mathbb{1}(y_{\text{i}}=j) \frac{e^{z_i/T_t}}{\sum_{j=1}^{K} e^{z_j/T_t}} .
\label{teacher-loss}
\end{equation}

Unlike the case of the temperature of the teacher model, the temperature we utilize for training the student is very low where the loss function to train the student model is defined as   
\begin{equation}
 \mathcal{L}_{\text{student}}(T_s)= -\frac{1}{N} \sum_{j=1}^{N}  \mathbb{1}(y_{\text{i}}=j) \frac{e^{z_i/T_s}}{\sum_{j=1}^{K} e^{z_j/T_s}}, 
\label{student-loss}
\end{equation}
where $T_s$ is the temperature which is a very low value in the case of our student loss, $T_s$ = 1 .

For the distillation process meaning transferring the knowledge of the teacher to the student, we are going to combine two different losses, one is the student loss and the other is the  Kullback-Leibler (KL) divergence or cross-entropy loss. The student loss computes the difference between the student predictions and the ground-truth labels, while the other computes the difference between soft student predictions and soft teacher predictions by using the Kullback-Leibler (KL) divergence or cross-entropy loss between the teacher's and student's probability distributions. The KL divergence measures the difference between two probability distributions, here the idea is to minimize this difference during the distillation process to the point that the knowledge from the teacher has successfully been distilled to the student. The KL divergence loss can be defined as
\vspace{-0.05in}
\[\mathcal{L}_{\text{KL}} = D_{\text{KL}}(P_{\text{teacher}}(i,T_t) || P_{\text{student}}(i,T_s))\]
\vspace{-0.1in}
\begin{equation}
=\sum_i P_{\text{teacher}}(i,T_t) \log\left(\frac{P_{\text{teacher}}(i,T_t)}{P_{\text{student}}(i,T_s)}\right).
\label{distillation-loss}
\end{equation}

From Equation \ref{distillation-loss}, we can observe that the cross-entropy loss is calculated between the softmax probability outputs of the teacher and student models at a very low temperature $T_t$ and $T_s$ greater than 1. Then, the final loss $\mathcal{L}_{\text{distillation}}$ used for the distillation process is computed as a combination of both the student loss and the KL divergence loss and is given by
\vspace{-0.1in}
\begin{equation}
 \mathcal{L}_{\text{distillation}}= \alpha * \mathcal{L}_{\text{student}} + (1-\alpha) * \mathcal{L}_{\text{KL}},
\end{equation}
where $\alpha$ is a weight parameter between 0 and 1. For our performance evaluation, we select $\alpha = 0.1$ to give more importance to the distillation loss. 

\begin{figure*}[h!]
    \centering  \includegraphics[width=0.85\linewidth]{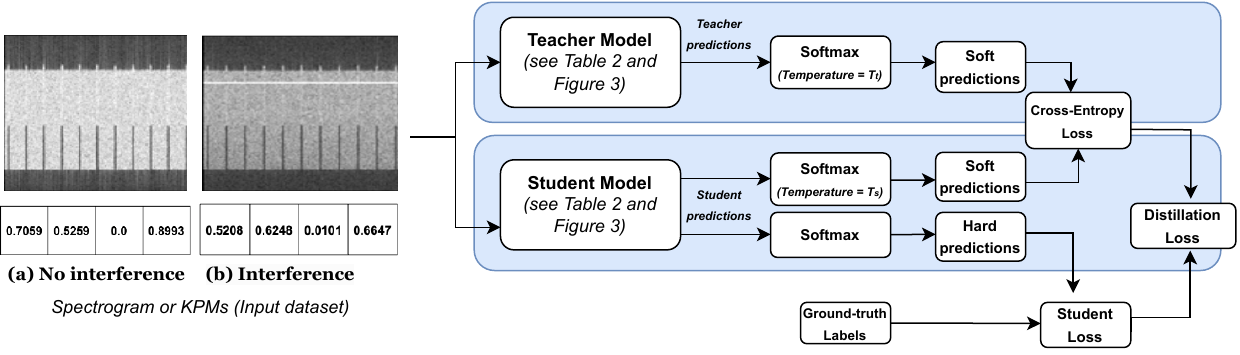}
    \vspace{-0.15in}
    \caption{Overview of the distillation-based defense.} 
    \label{distillation}
    \vspace{-0.2in}
\end{figure*}

Fig. \ref{distillation} shows an overview of the whole distillation technique explained so far. Both the teacher and student models are structured similarly to what has been defined in Table \ref{spec-model} and Fig. \ref{fig:kpm-model}. However, we have different loss functions for the teacher ($\mathcal{L}_{\text{teacher}}$) and the student ($\mathcal{L}_{\text{student}}$). These functions allow us to adjust the scaling of the logits with different temperatures where we use a high temperature for the teacher and lower temperature for the student.

\subsection{Baseline Defense: Adversarial training}
Adversarial training is a well-established technique for enhancing the robustness of models against adversarial attacks. Notably, in works such as \cite{goodfellow2014explaining} and \cite{madry2017towards}, researchers have addressed the issue of adversarial attacks by advocating for robust optimization techniques. In \cite{madry2017towards}, the authors presented a theoretical framework that formulates security against adversarial attacks as a saddle point problem, providing a principled approach to addressing this challenge. Adversarial training involves augmenting the training dataset with adversarial samples, and this method has demonstrated effectiveness in the field based on their research findings. The intuition behind this is that by having training data with some adversarial examples will help to improve the generalization capabilities of the model and also improve accuracy. To this end, we also employ adversarial training by augmenting the training samples with data generated from our attacks, specifically at an epsilon value of 0.02. 
\label{sec:secureoran}

\section{O-RAN Testbed and Dataset Generation}
\label{sec:testbed}
In this section, we present the implementation details of our OTA LTE/5G O-RAN testbed. Subsequently, we discuss how the testbed is utilized to generate both the spectrogram and KPM dataset for performance evaluations in Section 8.

\begin{figure}
    \centering  \includegraphics[width=3.3in, height=4cm]{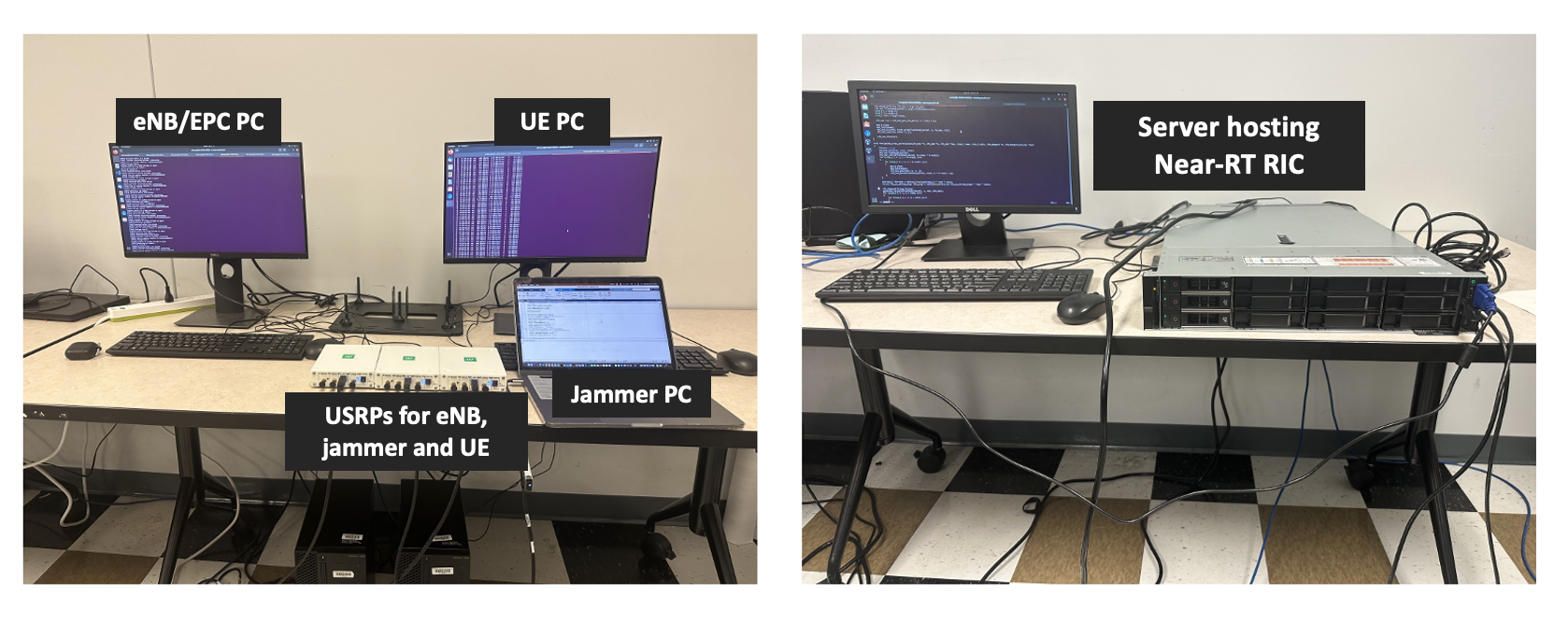}
    \vspace{-0.2in}
    \caption{O-RAN testbed. Left image shows RAN, UE, and the jammer USRPs. Right image shows the rack server hosting near-RT RIC that hosts InterClass xApp.}
    \label{fig:testbed}
    \vspace{-0.25in}
\end{figure}

\subsection{O-RAN Testbed Implementation}

In Fig. \ref{fig:testbed}, we provide an overview of our O-RAN testbed, which is composed of key elements including a RAN/Core network which are co-located on the same desktop, UE, jammer, and finally the near-RT RIC. 

The RAN/Core and UE are implemented using the open-source srsRAN cellular software stack (version 21.10), which is specifically designed for building LTE/5G cellular networks \cite{srsRAN}. Each of these are equipped with an ubuntu release 20.04 OS and running on an intel core i7-8700 having 6 CPU cores, 16GB RAM, 12 threads and running at a clock speed of 3.2GHz. The srsRAN software stack is a flexible and customizable platform for software-defined radio (SDR)-based RANs and UEs. To adapt srsRAN for the testbed's requirements, certain modifications were made to the codebase. These modifications include the creation of a buffer to store collected I/Q samples and the addition of specific RAN control capabilities such as switching between adaptive or fixed MCS. Both the RAN and UE are equipped with USRP B210 SDRs to handle the radio frequency (RF) front-end operations. SDRs are essential for the flexibility and programmability required in the O-RAN environment.

The near-RT RIC is hosted on a rack server and has the capacity to serve multiple RANs. The server hosting the near-RT RIC is an AMD EPYC™ 7443P with 24 CPU cores, 48 threads, 64GB RAM and a base clock speed of 2.85GHz. It acts as an intelligent controller for the RAN. The near-RT RIC interfaces with the RAN via an E2-lite interface, allowing it to make decisions and control RAN functions based on real-time data and network conditions. For the near-RT RIC, we developed lightweight internal components to facilitate our experiments and analysis, including:
\vspace{-0.1in}
  
    \smallskip \noindent $\bullet$ 
      \textbf{Lightweight E2 Interface (E2-lite)}: The interface we have adopted supports the essential closed-loop communication between the near-RT RIC and the RAN, playing a crucial role in the orchestration of O-RAN functions. Our interface design aligns with practicality and implementation efficiency, wherein we have opted for an E2-lite interface built upon the SCTP protocol. This approach closely parallels the functionality of the full E2 interface, enabling the exchange of control and report messages vital to the O-RAN system's operation. In the traditional E2 standard, RAN Functions are pivotal in defining service specifications and behavior facilitated through the E2 interface where RAN Functions inform the RIC about the capabilities supported by the RAN, laying the foundation for collaborative network operations. However, in the E2-lite interface, the process is streamlined, as it does not require explicit communication of RAN Functions. This simplification simplifies the connection setup and forgoes the need for subscription processes or inherent message differentiation between E2-lite nodes.
    \smallskip \noindent $\bullet$ 

    \textbf{Internal Messaging Infrastructure (IMI)/Policy controller}: This component enables coordination among various elements within the near-RT RIC. OSC implementation of this is the RIC message router. For our implementation, we develop a light weight version that is able to forward data from the RAN to the RIC database and is also able to relay control decisions from the xApps to the RAN. We implement the connections using the stream control transmission protocol (SCTP).
    
    \smallskip \noindent $\bullet$ \textbf{RIC Database and Shared Data Layer (SDL)}: As mentioned in Section \ref{dbsdl}, this serves as a repository for storing RAN information, which xApps utilize for making different inferences. Using the OSC documentation \cite{sdl}, we implement the RIC database which is redis-based and the SDL API for enabling multiple xApps to assess the database.
    

The jammer component is responsible for generating jamming signals that interfere with the uplink signal from the UE. These jamming signals are transmitted OTA and are an essential part of testing the network's resilience to adversarial interference. MATLAB is used to generate and transmit the jamming signals. Additionally, the O-RAN software community's open-source codebase is employed to implement the near-RT RIC. The source code for the near-RT RIC is compiled on the server to establish communication with the RAN.




In summary, the O-RAN testbed combines open-source software stacks, SDRs, and custom code modifications to create a flexible and programmable environment for testing and evaluating O-RAN network performance under various conditions, including adversarial interference generated by the jammer. The near-RT RIC plays a crucial role in network control and decision-making, and the system is designed to meet strict latency constraints, ranging from 10ms to 1s.

\vspace{-0.1in}
\subsection{Dataset Generation}
\label{datageneration}

\noindent{\textbf{Spectrogram data generation:}}
Our initial dataset comprises 10,000 spectrograms, which have been divided into two distinct classes for training purposes. Each spectrogram image is of the dimension (128,128,1) representing a gray scale image of width 128 pixels, height 128 pixels and 1 channel for intensity (grayscale). The first class, consisting of 5,000 spectrograms, represents the uplink UE SOI with no interference. These SOI spectrograms are derived from data transmitted at an uplink carrier frequency of 2.56GHz. To generate these spectrograms, we used 25 physical resource blocks (PRBs), which correspond to approximately 5MHz of bandwidth, necessitating a sampling rate of 7.68 Mega samples per second. We also generated uplink TCP traffic at a rate of 5MHz between the UE and the RAN using iperf3. We set up an iperf3 server at the RAN end, with the iperf3 client running on the UE.


The second class represents scenarios with interference, specifically continuous wave interference (CWI). These interference signals were generated at various gain values ranging from 30dB to 40dB. It is important to note that these gain values are relatively high, primarily aimed at assessing the vulnerability of intelligent components within the near-RT RIC to detect and handle interference. There are 5,000 CWI spectrograms in total.

The SOI signals are transmitted OTA using the open-source srsRAN stack, while the jamming signals are transmitted OTA on the same carrier frequency as the SOI. We use a MATLAB-generated script to generate the baseband I/Q samples for CWI signals and then utilize another USRP to transmit these signals OTA. Following the dataset generation, we perform various data processing steps such as resizing the image and converting to grayscale to prepare for model training. In Fig. \ref{distillation}, we can see the processed spectrograms belonging to the two classes we have considered and used as input dataset for the distillation procedure. The x-axis represents the time duration of each spectrogram. In our case, this is 10ms representing a LTE frame. The y-axis represents the frequency which tells us the amount of bandwidth occupied by the signal. In our case we used approximately 5MHz of bandwidth.


\noindent \textbf{\textbf{KPMs dataset generation:}} For the KPMs dataset, we use a total of 25,286 KPMs, consisting of 15,032 KPMs for scenarios with no interference (SOI) and 10,254 KPMs for cases with interference.

We employ the same experimental setup and parameters as used for generating the spectrogram dataset. However, in this case, we utilize KPMs obtained from the network to assess the presence of a jammer. Our primary focus here is on uplink metrics that could provide insights into the presence or absence of interference. Notably, we consider metrics such as uplink Signal-to-Interference-and-Noise Ratio (SINR), network bitrate/throughput, Block Error Rate (BLER), and Modulation and Coding Scheme (MCS). These metrics are selected based on their ability to effectively represent the presence or absence of an interference signal, supported by domain knowledge and experimental observations.

The KPMs are collected at varying intervals, ranging from every 0.5 seconds to 1 second. Similar to the spectrogram dataset, we perform several data processing steps to prepare the KPM dataset for training our models. Similarly, in Fig. \ref{distillation}, we can see the normalized and processed KPMs belonging to the two classes we have considered and used as input dataset for the distillation procedure. Here, m=4 is the number of metrics we consider and the metrics include uplink SINR, bitrate, BLER, and MCS, while t=1 indicating just one time window. All generated dataset used for the experiments can be found in \footnote[1]{\url{https://www.nextgwirelesslab.org/datasets}}.

\section{Performance Evaluation}
In this section, we evaluate the impact of the adversarial attacks and the distillation defense techniques on both InterClass-Spec and InterClass-KPM xApps in terms of ML model performance, network performance, and the overall O-RAN round-trip time (RTT).
\vspace{-0.1in}
\subsection{ML Evaluation}
We present the evaluation results of our robust ML models, considering the defense strategies outlined in Section \ref{defenses}. The performance metric used for ML model evaluation is accuracy, defined as:
$$Accuracy = \frac{TP + TN}{TP + TN + FP + FN}.$$

\begin{figure}
    \centering
    \includegraphics[width=2.7 in, height=1.9 in, trim={0 0.08cm 1.5cm 1.5cm},clip]{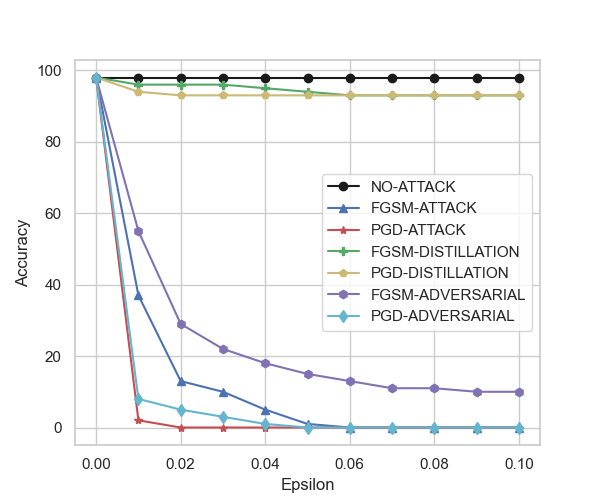}
    \vspace{-0.1in}
    \caption{Comparison of InterClass-Spec xApp accuracy vs epsilon for different attacks and defenses.}
    \label{subfig:accuracy_epsilonn}
    \vspace{-0.2in}
\end{figure}

\begin{figure}
    \centering
    \includegraphics[width=2.7 in, height=1.9 in, trim={0 0.08cm 1.5cm 1.25cm},clip]{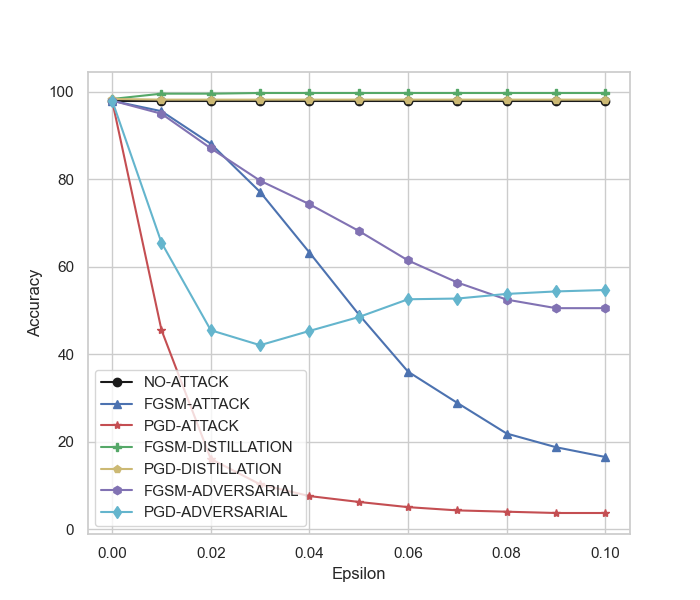}
    \vspace{-0.1in}
    \caption{Comparison of InterClass-KPM xApp accuracy vs epsilon for different attacks and defenses.}
    \label{subfig:accuracy_epsilon(kpms)}
    \vspace{-0.2in}
\end{figure}

Fig. \ref{subfig:accuracy_epsilonn} and Fig. \ref{subfig:accuracy_epsilon(kpms)} provide a comparative analysis of the scenarios involving no attacks on the models, different adversarial attacks, and the two defense mechanisms we have explored. For the InterClass-Spec model, as illustrated in Fig. \ref{subfig:accuracy_epsilonn}, the results show that adversarial training does not yield a substantial improvement compared to the distillation approach. The distillation method achieves accuracy comparable to those observed in the absence of attacks. On the other hand, the InterClass-KPMs model depicted in Fig. \ref{subfig:accuracy_epsilon(kpms)}, demonstrates that adversarial training results in some level of improvement, although it does not match the performance achieved through distillation.

Across both model types, the distillation approach consistently proves to be the most effective, achieving accuracy akin to those observed in attack-free scenarios for all epsilon values considered in the experiments. Similar results can be seen in \cite{7546524} and \cite{ozgur2022defensive} for distillation defense technique which achieve high performance.


It should be noted also that the adversarial attacks have more impact on the spectrograms compared to the KPMs. This can be attributed to the fact that the spectrograms data has a larger dimension size compared to KPMs. Larger spatial dimensions provide more room for perturbations to be added without significantly altering the overall appearance of the input. This can make it easier for adversaries to craft adversarial samples that can easily fool the models. On the other hand, smaller dimensions can make it more challenging to craft effective adversarial samples. 



\begin{figure*}[h!]
\begin{center}
\subfigure[Throughput (Spectrograms)\label{bitrate_cdf}]{
\epsfig{figure=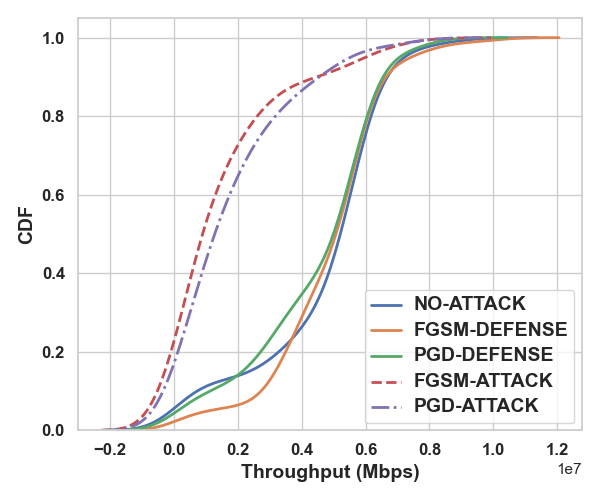,width=1.65 in}}
\subfigure[Throughput (KPMs)\label{bratekpmscdf}]{
\epsfig{figure=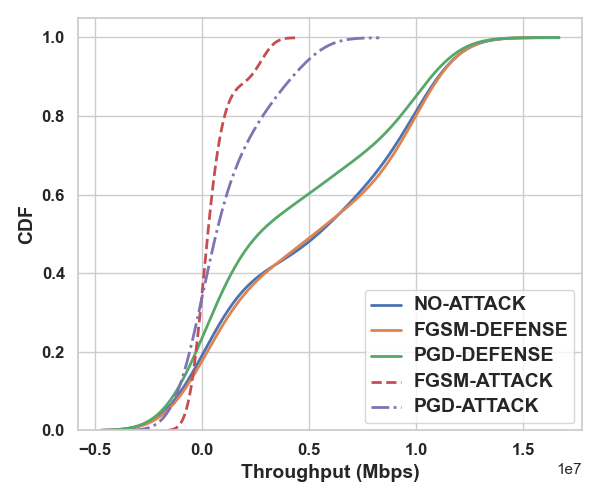,width=1.65 in}}
\subfigure[BLER (Spectrograms)\label{bler_cdf}]{
\epsfig{figure=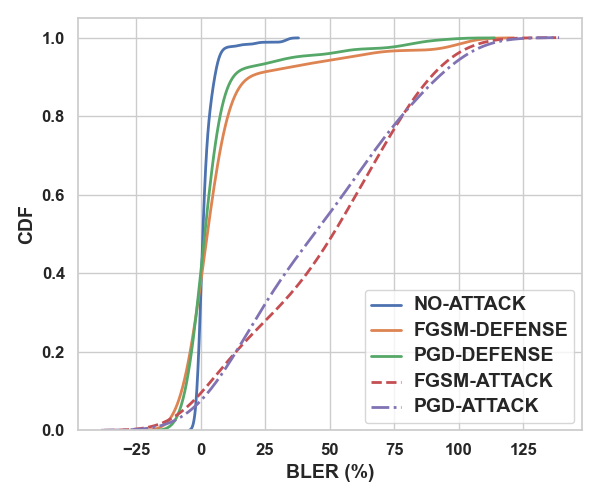,width=1.65 in}}
\subfigure[BLER (KPMs)\label{blerkpmscdf}]{
\epsfig{figure=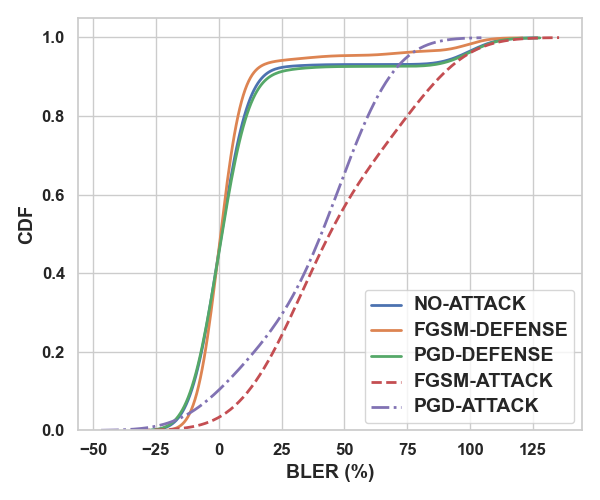,width=1.65 in}}
\vspace{-0.2in}
\caption{CDF plots of Spectrograms and KPMs models for various scenarios (no attack, under attack, and after defense) for throughput and BLER. } \label{spec-network}
\end{center}
\vspace*{-0.2in}
\end{figure*}

\subsection{Network Performance Evaluation}
\label{sec:network}
We present the results of our experiments using an attack budget of 0.1 with our OTA test-bed setup, as previously described. We evaluate the network performance under different conditions: no-attack, adversarial attacks, and the distillation defense technique. Specifically, we employ the PGD attack, performing a five-step gradient-based attack. To assess network performance, we initiate uplink traffic from the UE to the RAN, running for a total of 180 seconds. During the initial 90 seconds, the UE sends uplink traffic without any interference from the jammer. In the subsequent 90 seconds, we introduce 40dB OTA interference from the jammer.

From Fig. \ref{bitrate_cdf} and \ref{bratekpmscdf}, we can observe the cummulative distribution function (CDF) of the uplink throughput. We clearly see the impact of these adversarial attacks on these models in test time regardless of the closed loop latency constraints of the near-RT RIC to the RAN of between 10ms to 1s. From observation, the throughput of the network is clearly degraded under the two adversarial attacks considered and we can see that FGSM degrades the network performance slightly more than the PGD attack. This can be attributed to the fact that regardless of the FGSM attack being a one-step attack compared to PGD which we have used five steps, it will take slightly more time for the PGD to alter the most recent data in the database compared to the FGSM attack. From the plots, we can see that by using the defended models, we are able to achieve throughput performances comparable to the case of no-attack.

We can also observe the impact of the adversarial attacks on the uplink BLER in Fig. \ref{bler_cdf} and Fig. \ref{blerkpmscdf}. Similarly, we can see how these attacks increase the BLER of the UE due to inability to classify properly and we can also see an improvement on the BLER when we utilize the defensive distillation model. This improvement is relative to the scenario of no-attack.
\vspace{-0.15in}

\subsection{O-RAN Timing Evaluation}
To evaluate the performance of our O-RAN system, we have conducted an analysis of the overall timing of various components and steps in our system, particularly when using either spectrograms or KPMs as input data.

For the InterClass-Spec xApp case, we observe that the total RTT for the entire process is approximately 545.7ms, which is within the 1s latency requirement for near-RT RIC operation. During this process, a significant portion, specifically 49.84\%, of the time is spent on receiving I/Q samples from the RAN. In our case, we are collecting the last 10ms of I/Q samples from a 10ms LTE frame. This corresponds to 76800 I/Q samples, which equates to 614400 bytes of data. Each I/Q sample consists of 8 bytes. The data is then transmitted over the SCTP connection. The second most time-consuming operation is forwarding these bytes to the data processing microservice for converting to spectrogram, accounting for approximately 24.31\% of the total time. Other operational steps have also been considered in this analysis, contributing to a total RTT in Table \ref{rtt}. 

For the InterClass-KPMs xApp, the RTT, as observed in the analysis, is a total of 55.8ms, again comfortably below the 1s latency requirement. In this scenario, the time taken to receive the KPMs is approximately 52.39ms. This reduced time can be attributed to the fact that when using KPMs, we are sending only 4 features, amounting to a total of 20 bytes. Each KPM is cast as an integer before transmission over SCTP to the near-RT RIC.

In summary, comparing the RTT for both cases, we can conclude that using KPMs results in a significantly lower RTT when compared to using spectrograms. 
\vspace{-0.15in}
\subsection{Limitations and Future work}
As outlined in Section 7.2, we utilize high gain values ranging from 30dB to 40dB for our jamming signals to allow significant network performance degradation when misclassifying the high interference. However, further investigation and research are required to enhance the detection capabilities in scenarios characterized by low signal-to-noise ratio (SNR) or lower gain settings of the jammer especially in situations where network performance degradation is not the major concern but also identifying the presence of malicious transmitters or eavesdroppers.

Furthermore, our exploration in the wireless context has been limited to only two adversarial attack and defense techniques. There is a scope for extensive comparative analysis involving other prominent adversarial attacks, such as Limited-memory Broyden-Fletcher-Goldfarb-Shanno (L-BFGS), Jacobian-based Saliency Map Attack (JSMA), and Carlini and Wagner's C\&W. Additionally, there is potential for exploring detection methods like Distillation + Bayesian Uncertainty and Local Intrinsic Dimensionality.



\begin{table}
\caption{Break down of system timing of overall processes}
\vspace{-0.1in}
\begin{center}
\scriptsize
\begin{tabular}{|p{1.2in}|p{0.8in}|p{0.85in}|}
\hline
\textbf{Step} & \textbf{InterClass-Spec xApp} & \textbf{InterClass-KPMs xApp} \\
\hline
Receive I/Q samples or KPMs from the RAN & \textbf{272.0ms} & \textbf{52.39ms}\\
\hline
Time to forward data for processing& \textbf{132.68ms} & \textbf{22.82$\mu$s}\\
\hline
Data processing and storing in RIC database & \textbf{97.02ms} & \textbf{1.07ms}\\
\hline
Model inference & \textbf{44.03ms} & \textbf{2.35ms}\\
\hline
Control decision to RAN & \textbf{33$\mu$s} & \textbf{13$\mu$s}\\
\hline

Total time & \textbf{545.7ms} & \textbf{55.84ms}\\
\hline
\end{tabular}
\vspace{-0.15in}
\label{rtt}
\end{center}

\end{table}

\vspace{-0.1in}
\section{Conclusion}
In this paper, we conducted the first system-level analysis of adversarial attacks and defensive mechanisms targeting the intelligent components integrated as xApps within the near-RT of an O-RAN system. This analysis was carried out under the stringent latency constraints, ranging from 10ms to 1s. Our approach began with a thorough review of existing research that served as the foundation for this investigation, particularly focusing on simulations. Subsequently, we delved into the diverse types of data that can be stored in the RIC database, recognizing the key role this data plays in the wireless and O-RAN ecosystem. We then proceeded to conduct a systematic evaluation by creating two distinct ML models, each integrated into an xApp. Leveraging our in-lab O-RAN testbed, we rigorously assessed the impact of identified adversarial attacks and the corresponding defense mechanism on the overall network performance. Our findings identified the vulnerability of ML models to adversarial attacks and the subsequent adverse effects on both the models and network performance. Importantly, we demonstrated that these effects persisted despite the system's rapid RTT capabilities to meet stringent latency constraints. Our analysis of the two types of data also revealed the advantages and disadvantages of each in terms of RTT speed, providing valuable insights for future O-RAN system design and security considerations.

\vspace{-0.1in}
\section{Acknowledgement}
Authors acknowledge the funding support from NSF CCRI \#2120411, CNS \#2112471  and the Commonwealth Cyber Initiative (CCI).
\newpage


\end{document}